\documentclass[]{spie}
\usepackage[]{graphicx}
\usepackage{url}

\title{Problems with twilight/supersky flat-field for \\
wide-field robotic telescopes and the solution}

\author{Peng Wei\supit{1}, Zhaohui Shang\supit{2,1}, Bin Ma\supit{1}, Cheng Zhao\supit{3}, \\
Yi Hu\supit{1}, Qiang Liu\supit{1}
\skiplinehalf
\supit{1}National Astronomical Observatories, Chinese Academy of Sciences, Beijing, China; \\
\supit{2}Tianjin Normal University, Tianjin, China; \\
\supit{3}Tsinghua University, Beijing, China;; \\
}

\authorinfo{Send correspondence to Zhaohui Shang: zshang@gmail.com}

\begin{document}
 \maketitle

\begin{abstract} 

Twilight/night sky images are often used for flat-fielding CCD images,
but the brightness gradient in twilight/night sky causes problems of
accurate flat-field correction in astronomical images for wide-field
telescopes.  Using data from the Antarctic Survey Telescope
(AST3), we found that when the sky brightness gradient is minimum and stable,
there is still a gradient of 1\% across AST3's field-of-view of 4.3
square degrees.  We tested various approaches to remove the varying
gradients in individual flat-field images.  Our final optimal method
can reduce the spatially dependent errors caused by the gradient
to the negligible level.  We also suggest a guideline of flat-fielding
using twilight/night sky images
for wide-field robotic autonomous telescopes.

\end{abstract}


\section{INTRODUCTION} \label{sec:intro}

For modern observational astronomy, flat-field correction is a
necessary and critical step in data processing, which directly affects
the accuracy of photometry.  A flat-field
correction aims to correct the non-uniform response of the optical
system and the detector, so the uniformity of the light source (a
screen illuminated by a lamp, or twilight/dark night sky) is crucial
for the effectiveness of the flat-field correction.  

On the one hand, for wide-field robotic autonomous telescopes, due to
the wide field-of-view, the gradient in the surface brightness of the
twilight/dark night sky becomes a problem. In addition, the difference of the gradient
between different flat-field images affects the combined master
image precision and leaves some large-scale structures in the
master flat-field.  These large-scale structures introduce systematic
errors into the photometry.  

On the other hand, obtaining a dome flat-field for robotic telescopes
automatically has a higher level of technology costs, due to its
special design (usually compact dome or no dome) and its highly automated
operation without human intervention.

A typical field-of-view of photometric telescopes mostly cover 5--10
arcminutes on the sky. The surface brightness distribution of the
twilight/night sky in such a small size is sufficient to provide
uniform illumination for flat-field correction. 
However, large sky surveys require large field-of-view for survey
efficiency and there have been many telescopes with a field-of-view
much larger than 1 square degree.  Therefore the flat-field correction
for wide-field telescopes must take into account the
non-uniform surface brightness of the twilight/night sky light 
seriously.  For the twilight sky, previous studies have
found that there is a brightness gradient of 2\%--5\% per degree under a
typical twilight sky observing condition\cite{chromey96,
freudling07}.  And several correlations have been found between this
gradient and observing time (i.e., solar altitude) as well as 
the relative position between
the sun and the field-of-view where telescope points\cite{tyson93, chromey96}.
Even the dark night sky which is conventionally used to test the
flat-field flatness also becomes non-uniform on a large scale (e.g., 1 sq.
degrees)\cite{zhou04}.  If the sky flat-field images with stochastic
gradients are combined to create a master flat-field or supersky flat, its
accuracy and flatness will decline, but there has been little study on
this issue for large field-of-view telescopes.  


This work is based on large amounts of sky flat-field data taken by
the Antarctic Survey Telescope (AST3).  AST3 consists of three
wide-field 0.5/0.68m modified Schmidt telescopes with f-ratio f/3.73\cite{cui2008, Yuan2010}.  These catadioptric telescopes were
designed for wide field-of-view (4.3 sq. degrees) survey with a
shorter tube than traditional Schmitt telescope.  They are equipped with a large
single-chip 10Kx10K CCD operated in frame transfer mode with an effective
exposure area of 10Kx5K and 16 readout channels.  On the focal plane with the plane scale of 1
arcsecond per pixel, this corresponds to 4.3 sq. degrees (1.47 deg x2.93
deg).
The thermoelectric cooling (TEC) module is used to cool the CCD chip
to below the environmental temperature at Dome A in the winter,
which is about $- 60 ^\circ C $ on average.

By analyzing the flat-field data from the AST3, we aim at 1)
investigating the uniformity of the sky brightness, the spatial distribution of the
sky brightness and its variation with time, 2) analyzing and comparing
various components in flat-field and 3) studying and optimizing the
flat-field acquisition strategies and the automatic observation
procedures.  

\section{OBSERVATION AND DATA} \label{sec:data}

In January 2012, the first AST3 (AST3-1) equipped with the SDSS
i$'$-band filter was installed at Dome A, the highest place of the Antarctic
Plateau\cite{Li2012}.  When the polar day ended on 15 March,
AST3-1 began to acquire images.  Unfortunately, it stopped working on
May 8th because of a problem in the power supply system.  During the 56
day observing period, we obtained more than 22,000 images,
including 2,451 sky flat-field images.

We select the images with sky brightness in the middle of the dynamic
range of the CCD, between 15,000 and 30,000 ADU. The lower limit is to
ensure high S/N and the upper limit is to avoid non-linear response of
the CCD near saturation.  We also discard the flat-field images taken
with high CCD temperatures above $ -40^\circ C$, because the
high-level dark current at these temperatures is hard to correct
perfectly.  The average CCD temperature at which the rest of the
images were taken is $ -60^\circ C$, which corresponds to a dark
current of only 1 ADU/s.  Then we also require all images to have a sky
flux larger than 100 ADU/s so that the dark current correction is
small and does not introduce much noise.

Finally, there are 906 flat-field images left for analysis and their
average sky brightness is 22,500 ADU.  

\section{Gradients in the Sky Brightness}
\label{sec:gradient}

\begin{figure} \begin{center} \begin{tabular}{c}
\includegraphics[height=10cm]{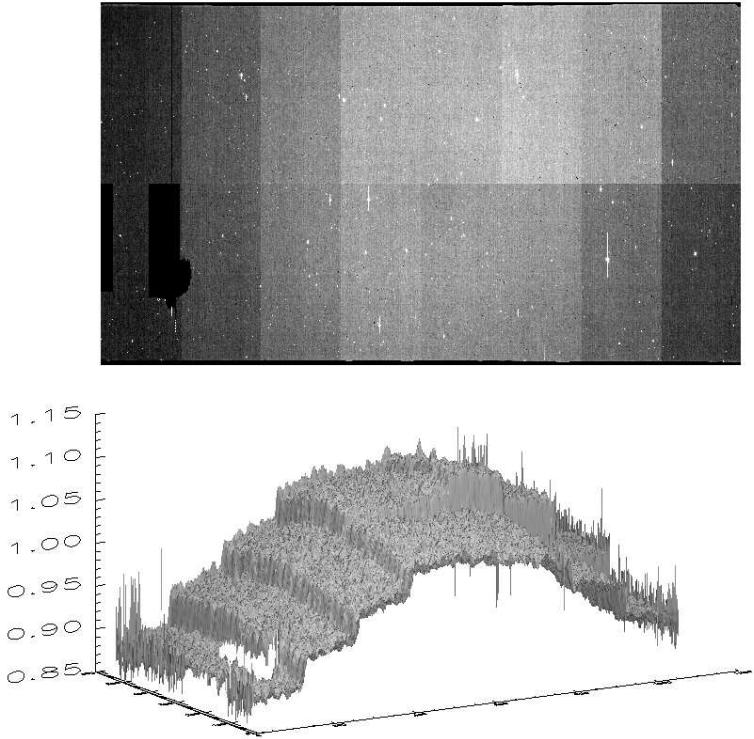} \end{tabular} \end{center}
\caption[image] {\label{fig:sp} The 2d and 3D plots of raw 
image a0425.810.fit, a flat-field frame taken by AST3-1.  The
black regions in lower-left part of the image are the bad pixels which were
masked.} \end{figure}

\begin{figure} \begin{center} \begin{tabular}{c}
\includegraphics[height=10cm]{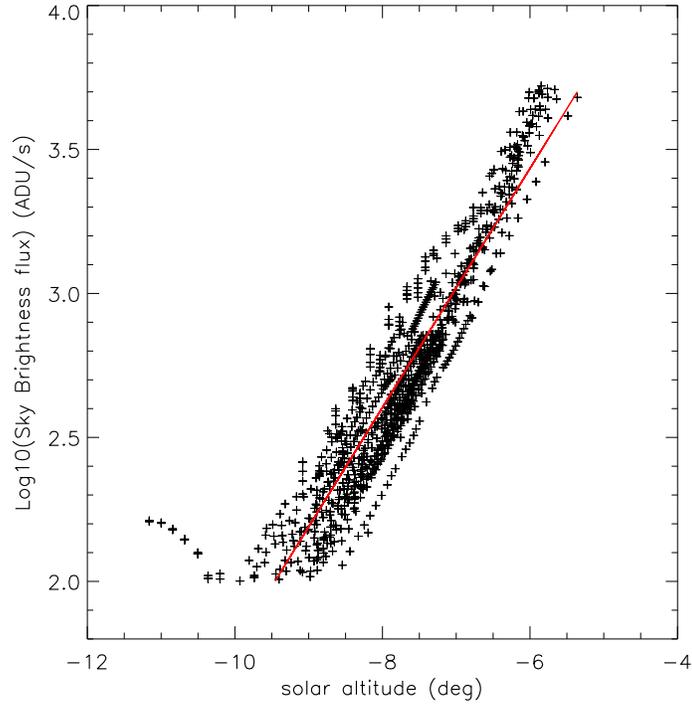} \end{tabular} \end{center}
\caption[alt] {\label{fig:alt} The correlation between the sky
brightness and solar altitude for each individual flat-field frame in
our sample.  The solid line indicates the best fit.  By checking
the images, we found that the weird tail with 
the solar altitude
below $-10^\circ$ are caused by the moonlight when the scattering 
sunlight did not dominate the sky
brightness during full moon.  } \end{figure}

\begin{figure} \begin{center} \begin{tabular}{c}
\includegraphics[height=10cm]{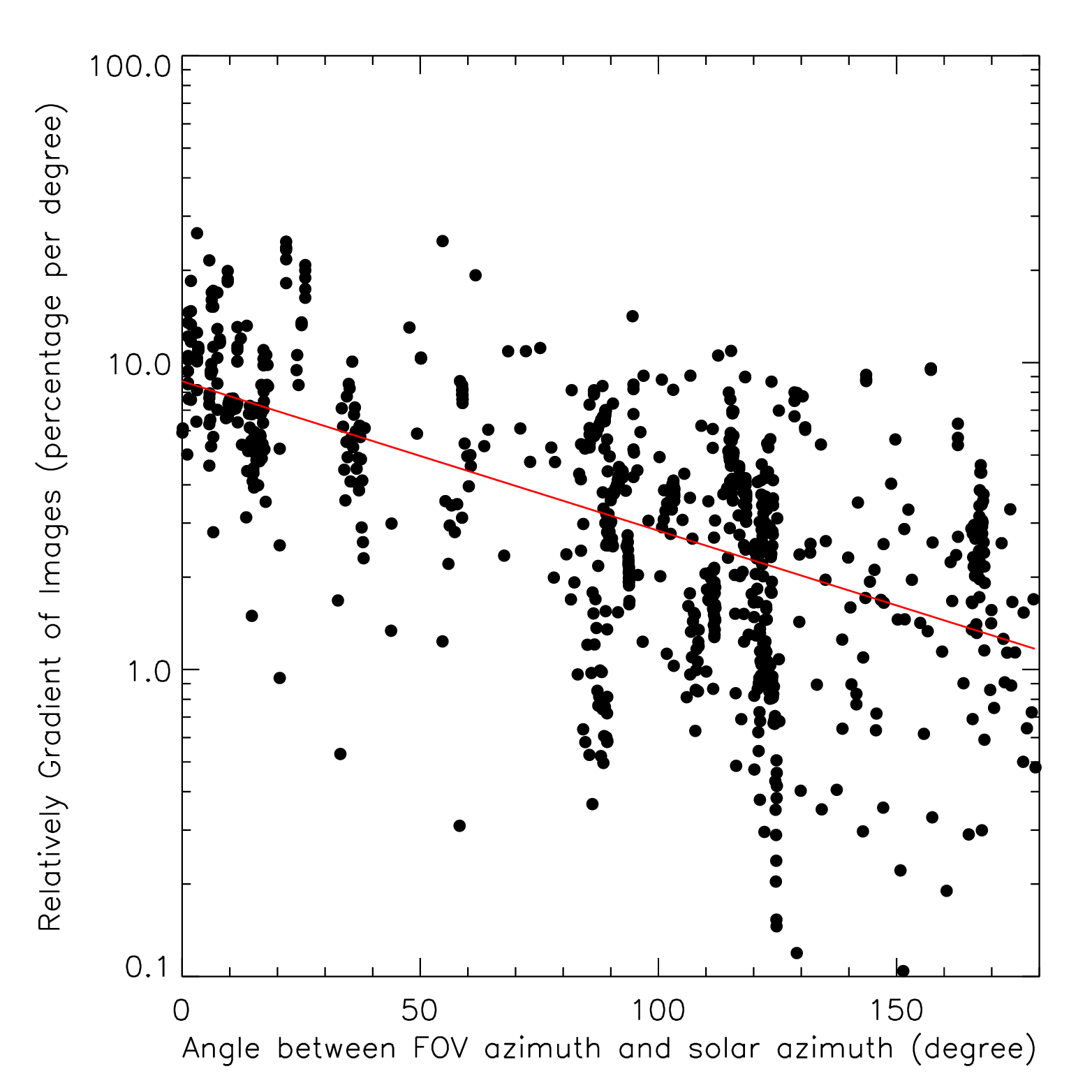} \end{tabular} \end{center}
\caption[azi] {\label{fig:azi} The gradient of sky brightness as a
function of the angle between telescope pointing azimuth and solar azimuth.
Every data point represents one individual flat-field frame in our
sample.  The solid line indicates the best fit.} \end{figure}

The original images of our sample have been pre-processed with a series of
steps, such as crosstalk correction (crosstalk phenomenon: the
multi-channel CCD readout inter-channel interference), overscan
correction, and dark current correction.  Finally, they are 
normalized and we have
many individual-exposure pre-processed flat-field image $F_i$.  

We take image a0425.810.fit as an example (Fig.~\ref{fig:sp}).  As we
can see, except for pixel-to-pixel nonuniform structure, there are
multiple large-scale structures in the individual flat-field image,
such as the vignetting and the obvious different mean values of each
readout channel due to different amplifier gains.  These all need to
be corrected by the final flat field correction.  However, in order to
derive the brightness gradient, we need to remove these large-scale
effects first, otherwise they can hide the brightness gradients and
make them hard to correct.  To do this, we apply an initial master
flat-field correction to individual images.

We combined those pre-processed images to obtain the initial master
flat-field by median stacking.  Then each individual flat-field image
is divided by this master flat-field. The resulted images consist
chiefly of the gradient component relative to the gradient of the
initial master flat-field.  Because of the median stacking
combination, the gradient in the initial master flat-field is small.
Also since we focus on correcting the pixel-to-pixel variation in the
flat field, the large-scale gradient in the initial master flat-field
does not affect our final results.
Then we fitted the individual resulted image with a two-dimensional inclined
plane $Z=a + bX +cY$ and derived the relative gradient
$G_i=\sqrt{b^2+c^2}$.

We check how the position of the Sun affects the twilight flat-field
images.  As expected, we found a close relationship between the sky
brightness $Flux_{sky}$ and solar altitude (see Fig.~\ref{fig:alt}).  
We fit this relationship and obtain an empirical function:

\begin{equation} \label{eq:alt} Flux_{sky} = 10^{0.415altsun + 5.926}
\end{equation} 

We also found that the sky brightness gradients of each frame $G_i$
is related to the azimuth angle $\theta_i$ between the telescope
pointing and the Sun (Fig.~\ref{fig:azi}).  The empirical function
from the fitting can be written as:

\begin{equation} \label{eq:azi}G_{i} =
10^{-0.00486\theta_{i} + 1.939} \end{equation}

These relationships confirm those in the previous studies about twilight sky
and the steady state ``standard model'' of Earth's
atmosphere\cite{tyson93, chromey96, patat06} .

\section{remove gradients and construct a master Flat-field}
\label{sec:method}

\begin{figure} \begin{center} \begin{tabular}{c}
\includegraphics[height=6cm]{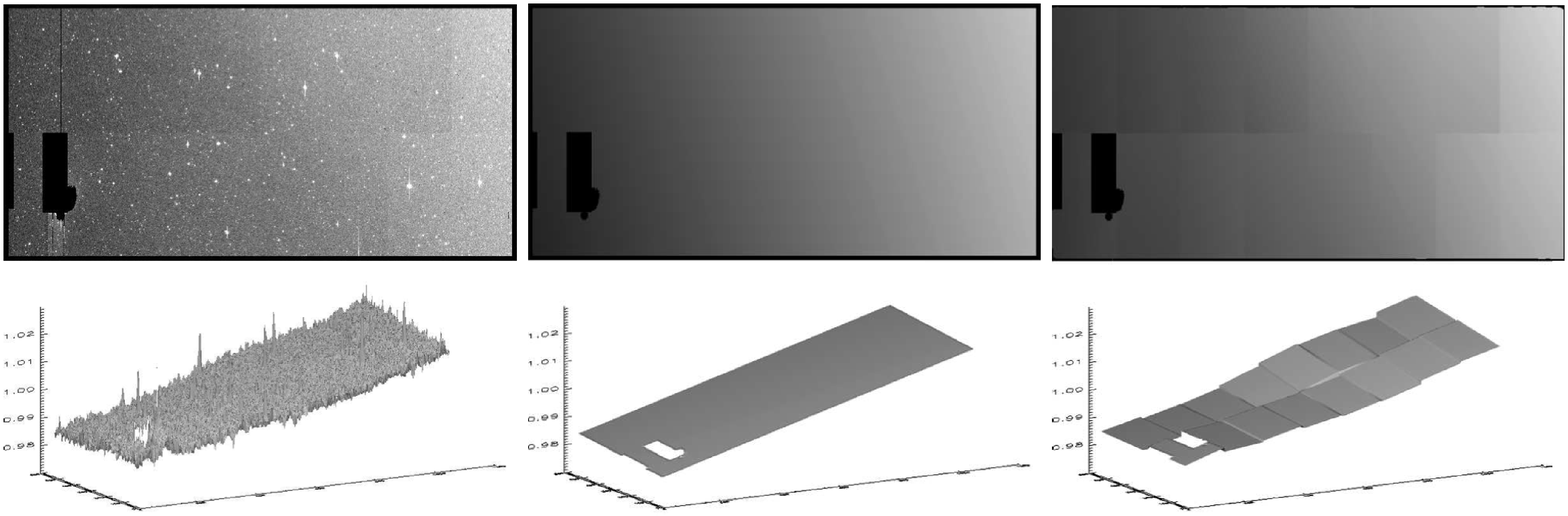} \end{tabular} \end{center}
\caption[res] {\label{fig:ex} Left: 2D and 3D plots of the flat-field 
corrected image 
of a0425.810.fit using the initial master flat; Middle:  
the full-frame fitting result of the left image;
Right:  the channel fitting result of the left image.
The black regions in the lower-left part of each
image are the bad pixels which were masked.  } \end{figure}

\begin{figure} \begin{center} \begin{tabular}{c}
\includegraphics[height=7.cm]{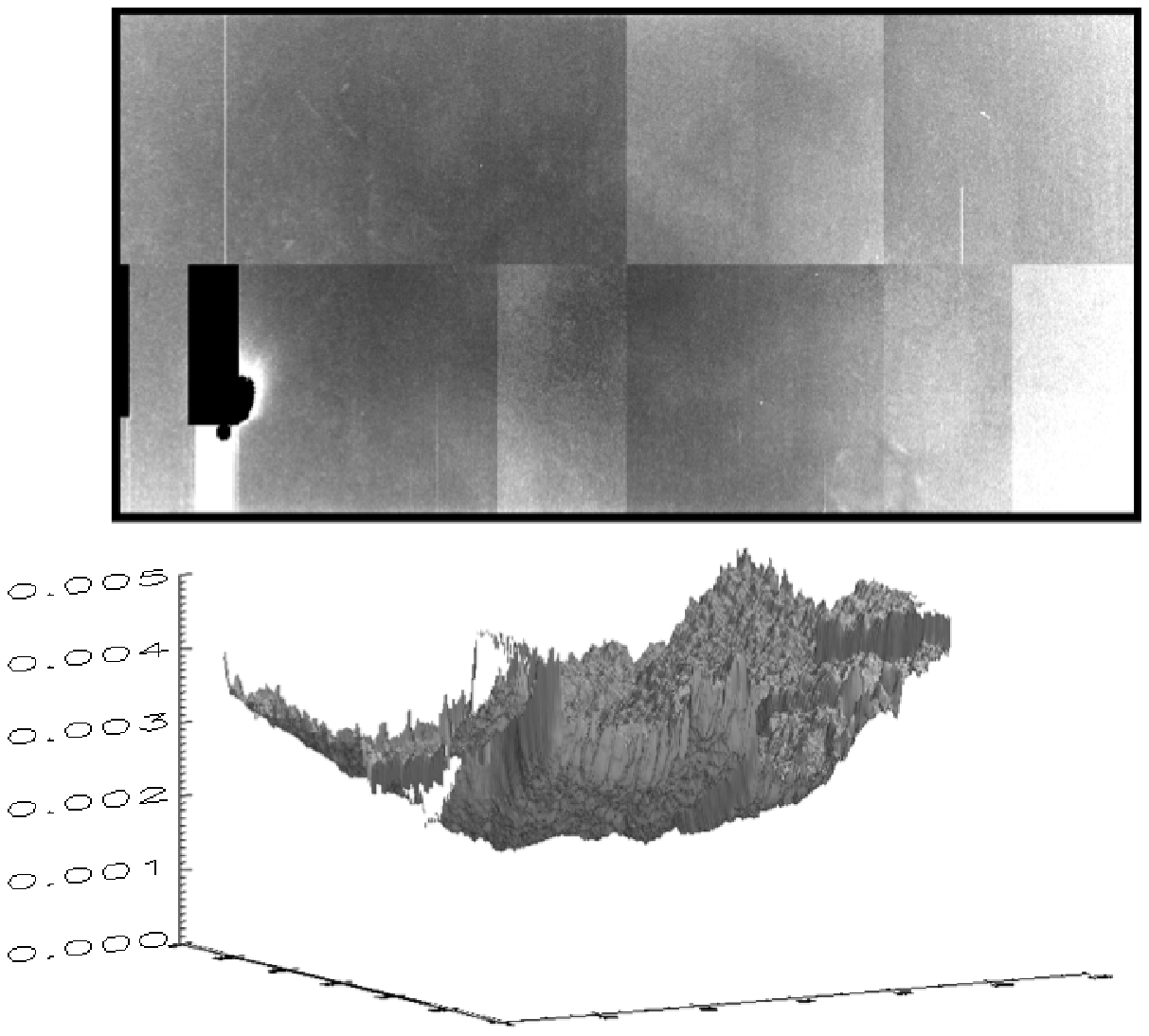}
\includegraphics[height=7.cm]{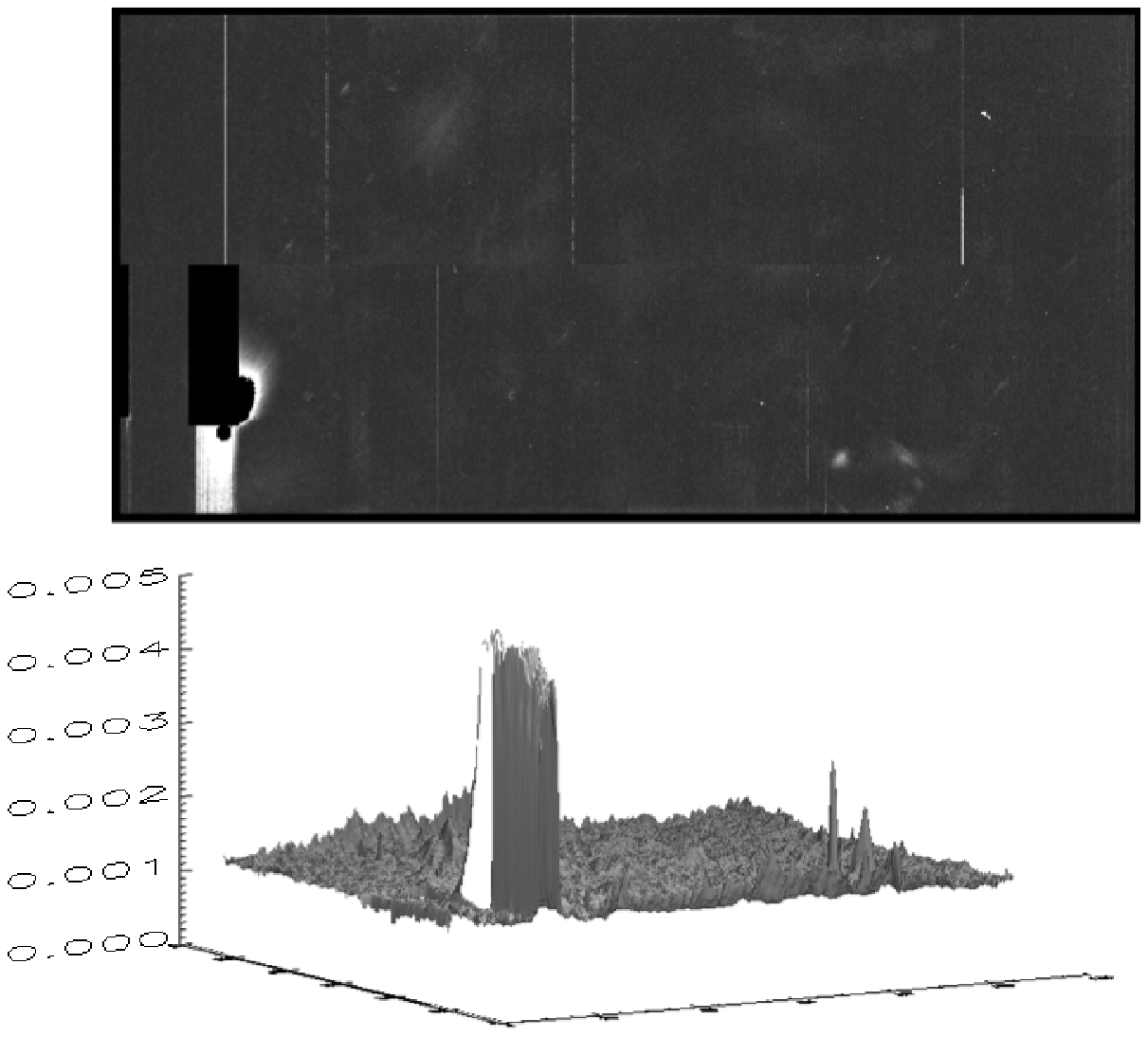} \end{tabular} \end{center}
\caption[rms] {\label{fig:rms} Left: 2D and 3D plots of the master
flat RMS
image without gradient correction.  Right: 2D and 3D plots
of the master flat RMS image with gradient correction (channel fitting). 
The black regions
in the lower-left part of each image are the bad pixels 
which were masked. The high
RMS pixels near those black regions are affected by electron
overflow from the bad pixels.  } \end{figure}

To build the final master flat-field, we have further rejected images
which have large gradient or large fitting residuals, because the vast
majority of them have interference structure due to the presence of
cirrus or scatter light.  We have also discarded the images containing
significant star-trailing due to telescope tracking problem during the
exposure.  The final sample has 200 images.

In order to minimize the spatially dependent photometric errors, we
need to fit and remove the gradients in these 200 individual
flat-field frames before combining them.  Previous
research\cite{chromey96} have found that the wide-field frames should
display a linear gradient in the sky background.  But by checking our
images, we found that the non-uniform structure is more complex than a
simple linear inclined plane.  In addition, there are some instant
Gain changes between channels due to the instability of the
amplifiers.  So instead of fitting the full frame, we choose to fit
each channel.  This method will introduce more degree of freedom for
fitting.  Eventually, to remove the gradient, we simply divide each image by its
fit.

In general, our method to remove the gradient in the flat-field images
and obtain the final master flat-field
can be expressed as:

\begin{equation} \label{eq:dualbri} Master Flat =
median\_comb\left(\frac{F_i}{
fitting(\frac{F_i}{median\_comb(F_i)})}\right).  \end{equation} 

\noindent
where $i$ denotes the ith image.  The $mediani\_comb()$ function means
the result from combining all images and the $fitting()$ function
means the fitting result and operates on each individual image.  The
fitting is actually applied to each channel of an image with a 2D plane
defined as $Z=a + bX + cZ$.  The parameter $a$ can reflect the channel
Gain variation relative to the initial master flat-field. Our fitting
method can correct this variation that otherwise introduces errors in
combination of images. 

We show an example of image a0425.810.fit in Fig.~\ref{fig:ex},
including the image after divided by the initial master flat-field,
fitting result with the full-frame fitting, and fitting result with
the channel fitting.  We note that the full-frame fitting does not
account for the extra change in the Gain of each channel.

To estimate the errors from this method, we also construct the RMS
image when combining the 200 individual flat-field images to the
master flat-field.  Fig~\ref{fig:rms} shows the RMS images with and
without the gradient correction.  It is clear that obtaining the
master flat-field after the gradient correction results in a much
smaller RMS.  We also note that the difference between full-frame
fitting and channel fitting is not big, the master flat-fields of both
have a final mean RMS of 0.1-0.2\%, but the channel fitting method can
reduce the spatial error in the combined master flat-field to the
negligible level compared to the photon noise.

\section{Strategy for Taking Flat-field} \label{secsstrategy}

Based on our experiments and results, we suggest the following
strategy for taking the sky flat-field images in order to achieve more
flatness and high accuracy for wide-field telescope.

1) Determine the start and end time of taking the flat-field images by
calculating the local altitude of the Sun;

2) Point the telescope toward the null point where the sky brightness
gradient is minimal (towards anti-sun direction, altitude 75 degrees); 

3) Take a short-exposure image to determine the zero point of the sky
brightness in eq.~\ref{eq:alt} for that date;

4) Calculate the exposure time based on eq.~\ref{eq:alt}, which
provides the relationship between the sky brightness and the solar
altitude.

5) Take flat-field images with tracking; 

6) After about 30 seconds of multiple exposures or a single exposure
longer than 30 seconds, point the telescope back toward the null point
again, and repeat steps 4 and 5.

Based on the above observing strategy, we have developed a program to
take flat-field images automatically.  The automated program improves
the efficiency and accuracy of flat-field taking during the limited
twilight time for each day.

\section{CONCLUSIONS} \label{sec:conclusion} 

We have studied the sky flat-field images of the wide-field telescope
AST3-1 from the observation in 2012 at Dome~A, Antarctica.  We have
developed the relationships between local solar position, telescope
pointing direction, and the flux and gradients of sky brightness.  We
have found that a 1\% level of the gradient within our field of view is
hard to avoid even if the telescope is pointed to the null point where
the gradient of sky brightness is minimum and stable.  This gradient
will have impacts on photometric accuracy and stability which are
especially critical to the time domain astronomy.  We tested various
fitting methods to remove the gradient in each individual flat-field
frames before combination.  The optimal method could reduce spatially
dependent errors in the combined master flat-field to the negligible
level.  

Based on the real-time sky brightness distribution, we have optimized
the twilight flat-field acquisition and suggested an automatic
flat-field observing procedure in order to  observe a more stable and
uniform sky.  With this observational strategy we can minimize the
nonuniformity of light source.  Our optimized flat-fielding not only
improves the image quality and photometric accuracy for achieving the
scientific goals of AST3, but will also provide scientific evidence
and guidance of flat-fielding for other wide-field robotic autonomous
telescopes.

\acknowledgments

This work has been supported by the National Basic Research Program of
China (973 Program) under grand No. 2013CB834900, the Chinese Polar
Environment Comprehensive Investigation \& Assessment Programmes under
grand No. CHINARE2014-02-03, and the National Natural Science
Foundation of China under grant No. 11003027, 11203039, and 11273019.

\bibliography{ref} \bibliographystyle{spiebib} \end{document}